\documentclass[12pt]{iopart}
\usepackage{graphicx}
\usepackage{hyperref}
\begin{document}

\title[On the genesis and nature of Palm Tree Modes in the JET tokamak]{On the genesis and nature of Palm Tree Modes in the JET tokamak}

\author{C. Maszl$^{1,2}$, V. Naulin$^3$, R. Schrittwieser$^2$ and JET EFDA Contributors$^\dag$}

\address{$^1$ Funding Support and Industry Relations, TU Wien, Austria\\
	$^2$ Institute for Ion Physics and Applied Physics, University of Innsbruck, Austria\\
$^3$ PPFE, Department of Physics, DTU, DK-2800 Kgs. Lyngby, Denmark\\
}
\ead{christian.maszl-kantner@tuwien.ac.at}
\begin{abstract}
Long-lived, highly localized structures called palm tree modes (PTM) are observed in the edge plasma of the JET tokamak. Although PTMs are well documented, little is known about the mechanisms which produce these structures. In the case of the PTM, an ELM-postcursor, its genesis is usually explained by ergodisation of the magnetic field due to edge localized modes and the appearance of a seed magnetic island which evolves into a PTM later.\\
In this study we try to invoke a creation mechanism based on the concepts and observations in edge plasma turbulence. An interesting aspect of plasma turbulence is the occurrence of coherent, long-lived structures in the scrape-off-layer (SOL). These localized and magnetic-field-aligned regions with higher or lower plasma densities are called blobs and holes. Measurements show that these filaments carry parallel currents. We thus here interpret ELM-filaments as massive blobs and the interspace between these filaments as holes.\\
We demonstrate that a forward-modelled closed current filament on a $q=3$ surface produces similar magnetic fluctuations as measured by the JET in-vessel magnetic pickup coils if a PTM is present. From that we deduce that if a hole is captured on a $q=3$ surface after an ELM-crash, a PTM equivalent signature is generated. If the ELM-filament itself is captured on a $q=4$ surface, a signature equivalent to an outer mode appears.
\end{abstract}

\maketitle
\section{Introduction}
\label{sec:introduction}
There is evidence for the existence of coherent long-lived highly localized structures in the edge of tokamak plasmas. These are found for example in the edge of JET as confined current ribbons or filaments (outer modes) \cite{solano2010} and palm tree modes (PTM) . Although these phenomena are well documented, little is known about the mechanisms which produce these structures. In the case of the PTM its genesis is usually explained by ergodisation of the magnetic field due to the ELM and the appearance of a seed magnetic island which evolves into a PTM \cite{koslowski2005}.\\
 Here we try to invoke a creation mechanism based on the concepts and observations of plasma turbulence. Turbulence in tokamaks is an active field of research and great efforts are spent, experimentally and theoretically, to quantify and understand particle, energy and momentum transport \cite{conway2008, Ionita2013}.  An interesting aspect of plasma turbulence is the occurrence of coherent, long-lived structures in the SOL. These localized and magnetic-field-aligned regions with higher or lower plasma densities are called blobs and holes, respectively.They lead to intermittent transport through the scrape off layer (SOL), which can effect  the integrity of plasma facing components. Blobs and holes were observed first by Zweben in 1985 during studies of edge density turbulence with an array of Langmuir probes in the Caltech research tokamak \cite{zweben1985}. In the poloidal plane, blobs and holes look like drifting quasiparticles. In the last years the origin of localized filamentary blobs has been identified as the edge shear layer, where zonal-flows shear off meso-scale coherent structures \cite{xu2009}. These filaments carry parallel currents \cite{vianello2009}. A significant amount of radial transport (up to $60\%$) might be carried by blobs \cite{bisai2005}.\\
ELM filaments, as observed with fast camera diagnostics \cite{scannel2007}, show certain similarities with blobs in plasma turbulence. Despite their size ELM filaments are likewise current carrying, magnetic-field-aligned coherent structures \cite{vianello2011} with higher densities and excess temperatures \cite{kamiya2007}, compared to the background plasma. According to \cite{myra2007}, ELMs should leave current holes behind because of total current conservation on short time scales in the torus. These observations lead us to the central hypothesis of this paper, that ELMs create current and density holes in the edge plasma and that their motion can lead to phenomena like PTMs.
\section{Experimental results}
\begin{figure}[ht]
	\centering
	\includegraphics[height=7cm]{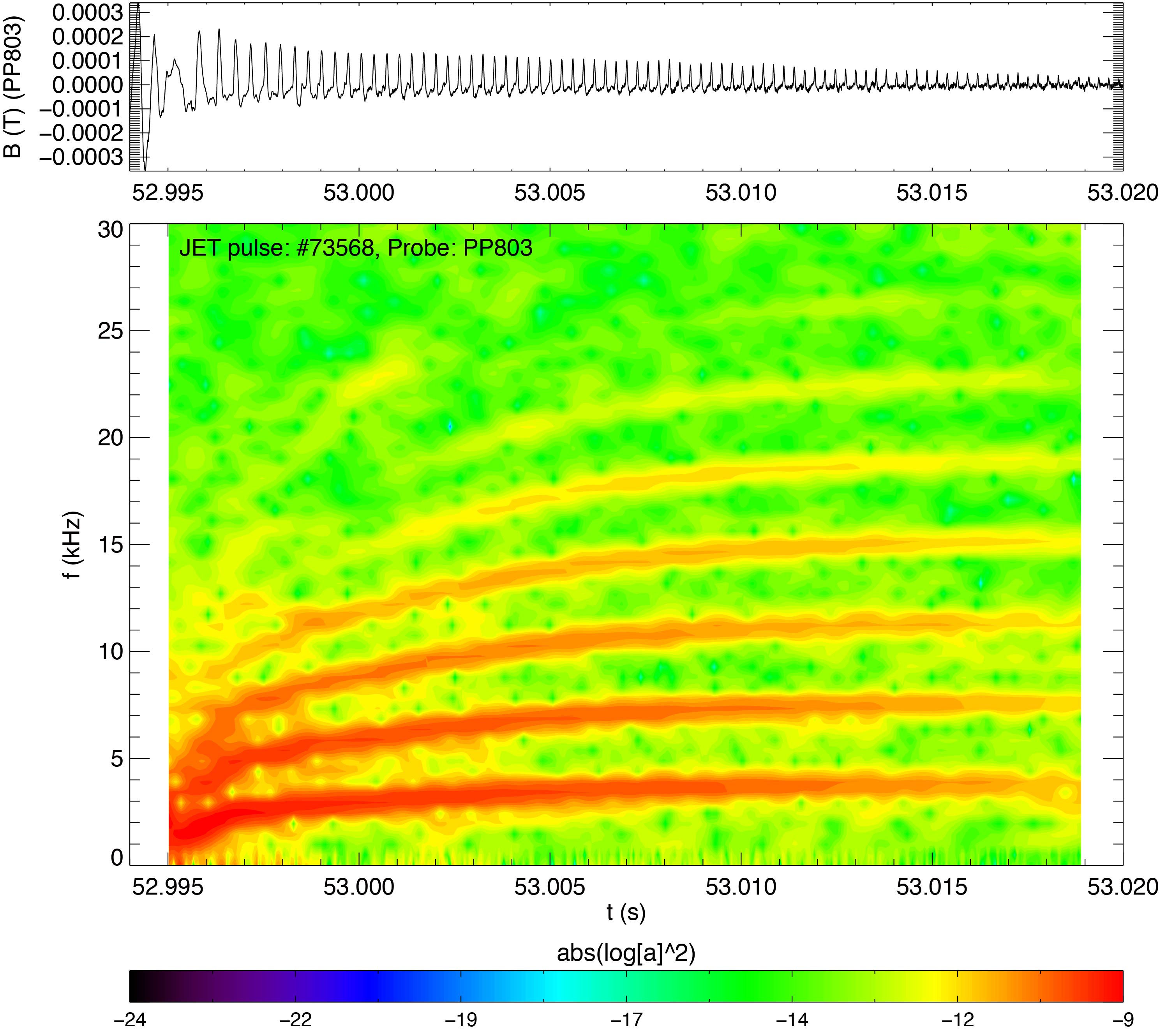}
	\caption{FFT of a PTM measured by coil PP803 from the poloidal limiter array (JET pulse: $73568$, $t_0=12.995$~s). The size of the FFT windows is $\tau=2$~ms and the time shift between the windows $dt=64$~$\mu$s.}
	\label{fig:ptmpaperfig01}
\end{figure}

The palm tree mode is an ELM post-cursor which was, as far as we know, only observed in JET. It occurs in a wide variety of plasma conditions, as long as the $q=3$ is located in the ELM-perturbed region. PTMs can be studied with the electron cyclotron emission (ECE), soft X-ray diagnostic (SXR) which covers the edge plasma and magnetic diagnostics at the limiters and the first wall \cite{koslowski2005}. Figure~\ref{fig:ptmpaperfig01} depicts a sliding window fast Fourier spectrum of a PTM magnetic signal and its characteristic appearance (JET pulse: \#73568, $t_0=12.995$~s).\\

The PTM starts after the ELM crash at $t_0=12.995$~s. Because of the strong magnetic perturbations of the ELM the actual onset time of the PTM cannot be deduced with certainty. Characteristic features are an increase in frequency for the first ms and the observation of rich harmonics, which indicate strong localization of the feature. These characteristics can also be observed in ECE and SXR signals and are the inspiration for the flowery name. The amplitude is decaying with time. Typical lifetimes of PTMs are around 23~ms although PTMs with lifetimes up to 60~ms can be observed. If an ELM is triggered during a PTM event, the PTM is terminated immediately. \\

JET has a number of in-vessel magnetic coils with sample frequencies in the range between one and two MHz. These coils and coil arrays were used to study the helical structure and localization of the PTM as well as growth and decay. The ECE diagnostic was used to study temperature perturbations. The combination of both diagnostics allowed a rough estimate of the location. Furthermore, charge exchange recombination spectroscopy (CXRS) was used to correlate the rotation frequencies of PTMs with the edge rotation. For this study a PTM database was created containing 36 PTMs. 24 of them are not superposed by other MHD activity (see Appendix Tab.~\ref{tab:ptm_db_I} and Tab.~\ref{tab:ptm_db_II} for more details). This is important because other MHD activity would modulate PTM signals and circumvent the study of growth and decay behaviour. PTMs were searched which are not stopped by a subsequent ELM, which limit the lifetime, to allow investigations on dissipative processes. For two PTMs the diagnostic coverage was more extensive (JET pulse: \#52011\footnote{Session Aims: Influence of triangularity on confinement, pedestal and SOL parameters} $t=19.14-19.22$~s  , \#73568\footnote{Session Aims: Scaling of confinement and pedestal with rho*, beta} $t=12.8–13.2$~s). Aside from fast magnetics also fast ECE with higher time resolution ($f_s=250$~kHz) was available. Therefore, these two pulses were studied in more detail. The JET control room physics summary is presented in Tab.~\ref{tab:jet_pulses}

\begin{table}[ht]
	\caption{JET Control Room Physics Summary for the pulses \#52011 and \#73568. $B_t$ toroidal magnetic field, $I_p$ plasma current, $n_e$ electron density, $T_e$ electron temperature, $P_{NBI}$ neutral beam injector power, EFCC error field correction coils, - not available/applicable}
	\begin{center}
		\begin{tabular}{|c|c|c|c|c|c|c|c|}
			\hline
			Pulse     & $B_t$(T)                 & $I_p$(MA) & $n_edl$($10^{19}$m$^{-2}$)          & $T_e$(keV)        & $P_{NBI}$(MW)        &	EFCC & Year		\\
			\hline
			52011	& 2.67	& 2.51	& 17.5	& -	  & 11.8 	& -	& 2000\\
			73568	& 2.00	& 2.20	& 12.4	& 4.6 & 10.7	& off & 2008\\ 
			\hline
		\end{tabular}
	\end{center}
\label{tab:jet_pulses}
\end{table}%
\subsection{Rotation}
\label{sec:Rotation}
\begin{figure}[ht]
	\centering
	\includegraphics[height=7.00cm]{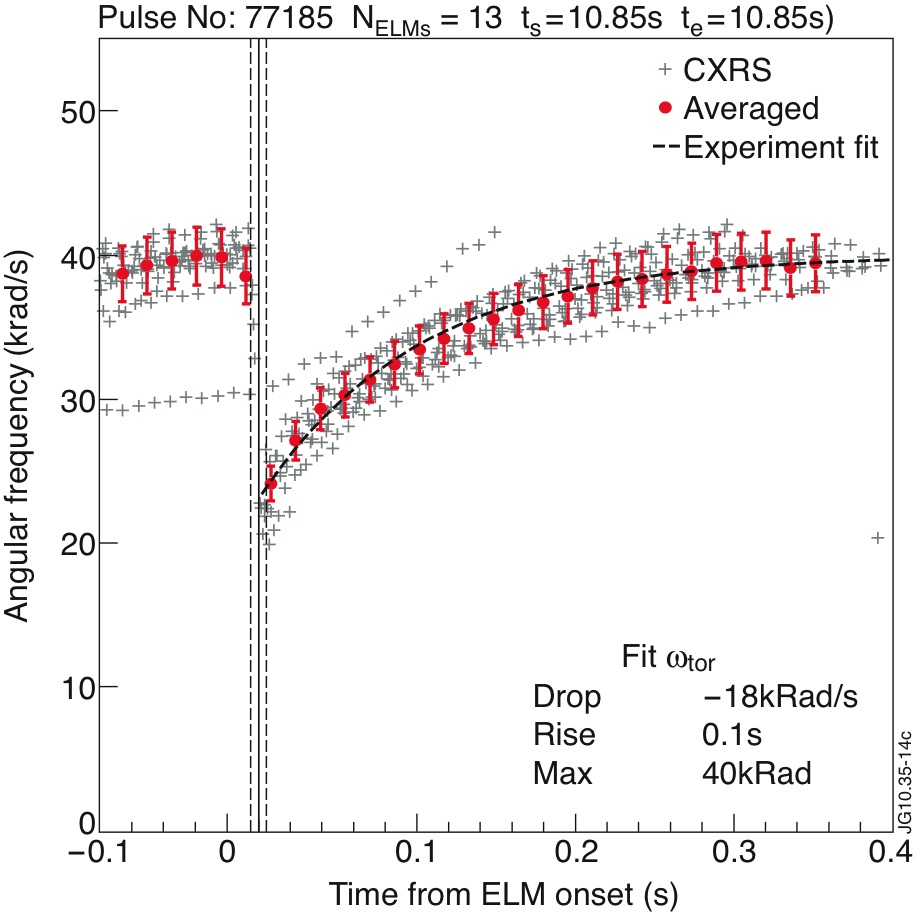}
	\caption{Edge rotation as measured by CXRS. Recovery after ELM induced momentum losses (vertical line) can be observed. Courtesy by Thijs W. Versloot.}
	\label{fig:ptmpaperfig02}
\end{figure}

Fourier spectra of PTMs show rich harmonics and also a frequency increase in the first ms. It is known that ELMs cause momentum losses in the pedestal region. These are usually studied with CXRS \cite{versloot2009}. The rotation in figure~\ref{fig:ptmpaperfig02} drops almost by $20$~krad/s after an ELM event.\\ 

\begin{figure}[ht]
	\centering
	\includegraphics[height=7.00cm]{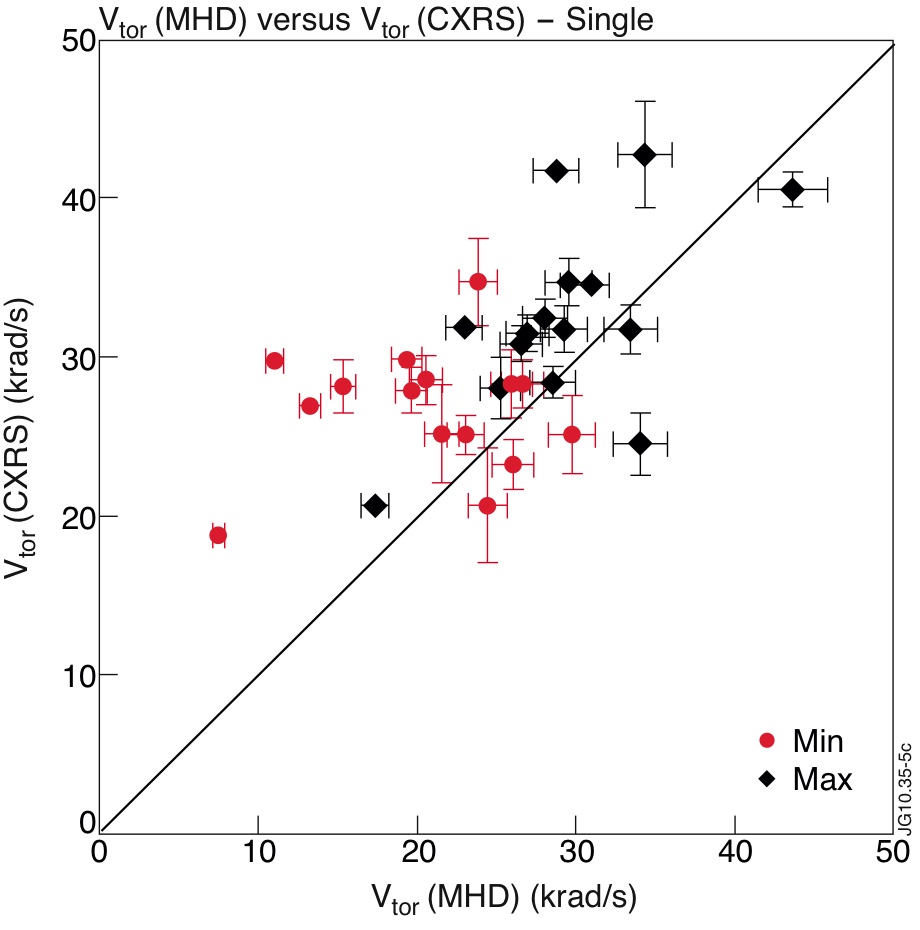}
	\caption{Comparison of initial (Min) and final (Max) velocities of PTM (MHD) and pedestal frequencies (CXRS) for single ELM data. The solid line indicate unity. Courtesy by Thijs W. Versloot.}
	\label{fig:ptmpaperfig03}
\end{figure}

It was found that PTMs are co-rotating with the edge plasma \cite{koslowski2005}. Furthermore the initial and final frequencies of PTMs and the pedestal are in the same order of magnitude as this co-rotation \cite{maszl2011}. A comparison is given in Fig.~\ref{fig:ptmpaperfig03}. Here the initial frequencies are colour-coded red and the final frequencies black. The solid line indicates unity.\\

The frequency increase and hence toroidal angular velocity $v_{tor}$ in rad/s of the PTM in JET pulse \#73568 at $t_0=12.994$~s (Fig.~\ref{fig:ptmpaperfig01}) can be fitted by 
\begin{equation}
v_{tor}=m_0+m_1\Big(1-e^{-m_2t}\Big).
\label{equ:ptmrot}
\end{equation}

with the fit parameters $m_0=14816.5$~rad/s, $m_1=9235.7$~rad/s and $m_2=189.4$~s$^{-1}$. The initial and final frequencies as seen in CXRS and MHD data are comparable. Surprisingly, the frequency increase extracted from MHD-data is approximately one order of magnitude faster. This discrepancy has to be addressed in future research.
\subsection{Temperature perturbations}
\begin{figure}[ht]
\centering
\includegraphics[height=8.00cm]{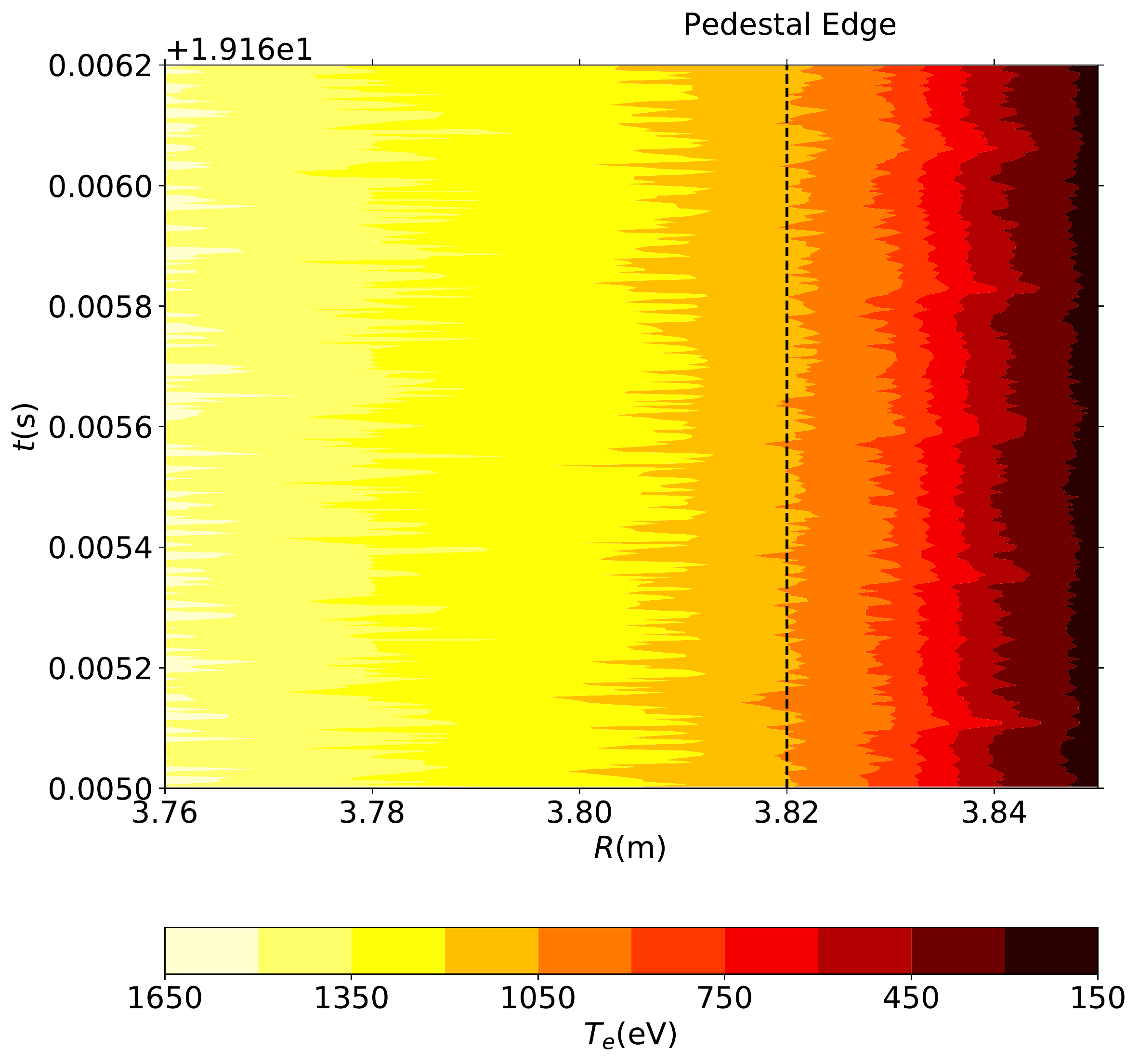}
\caption{The electron temperature $T_e$ as function of time $t$ and major radius $R$,  measured by ECE (KK3-CATS, JET pulse: $\#52011$). The temperature perturbations due to the PTM are visible in the edge region of the pedestal.}
\label{fig:ptmpaperfig04}
\end{figure}

PTMs can also be seen in the signals of the ECE diagnostic. Surface plots of ECE data locate the PTM in the gradient region of the pedestal (Fig.~\ref{fig:ptmpaperfig04}). However, ECE data have to be considered with care because they rely on equilibrium reconstruction. The exact spatial localisation of MHD modes is therefore not straightforward \cite{baruzzo2009}.\\

\begin{figure}[ht]
\centering
\includegraphics[height=14.00cm]{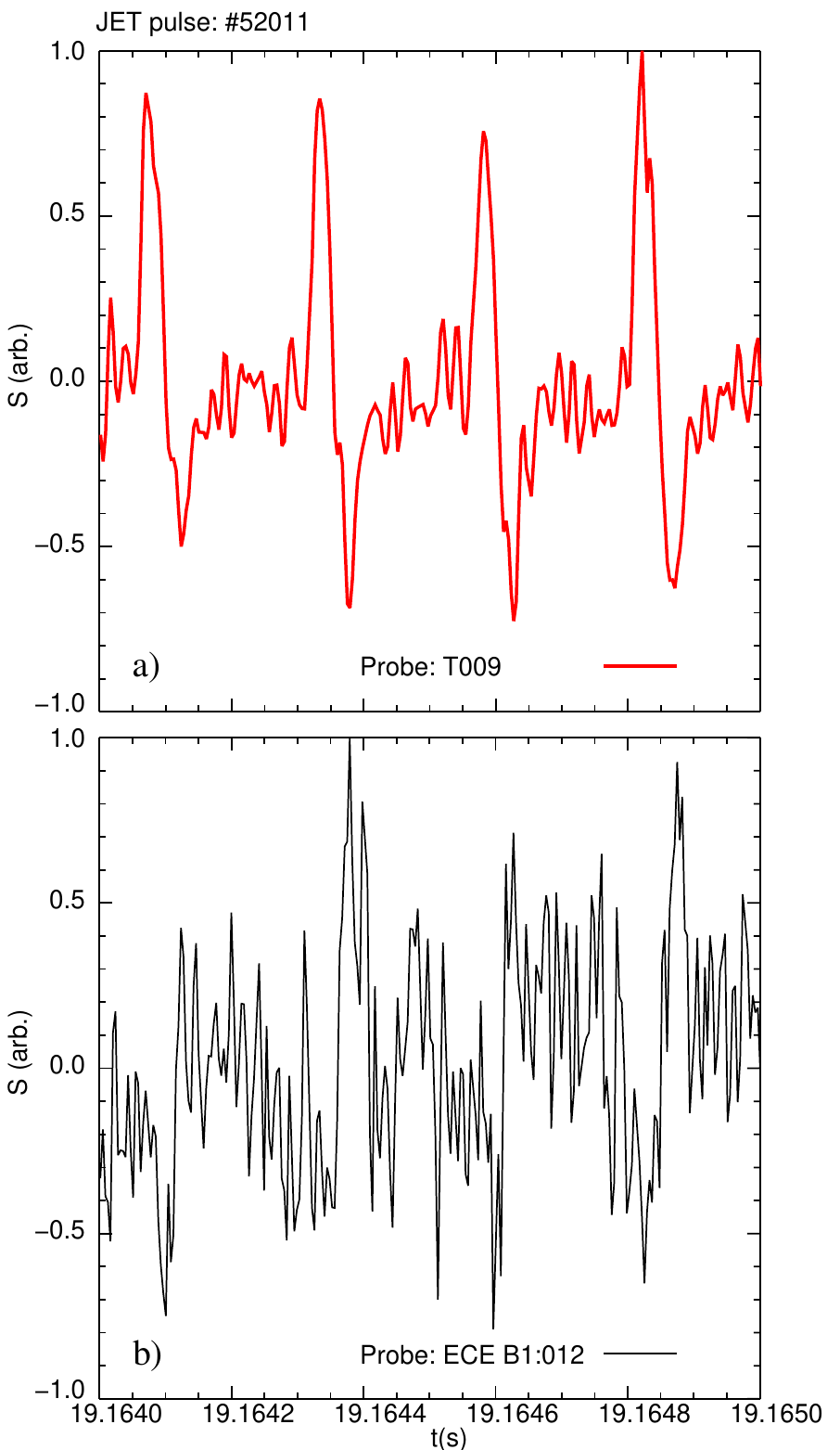}
\caption{Temporal comparison of a a) $\textrm{d} B/\textrm{d} t$ signal from magnetic coil T009 (red) in octant~7 (JET pulse: $\#52011$) and b) an electron temperature signal (ECE B1:012, black). Both signals $S$ are normalised to $1$.}
\label{fig:ptmpaperfig05}
\end{figure}

Moreover, the influence of the edge current density will alter the equilibrium in the plasma edge. PTMs are capable to disturb the equilibrium locally which is demonstrated in Fig.~\ref{fig:ptmpaperfig05}. There is a strong correlation and a $180^\circ$ phase shift between ECE channel B1:012 and coil T009 which is close to the line of sight of the diagnostic.\\

This is another direct evidence of a current perturbation due to the PTM. A similar observation was made by Koslowski et al. They found that the PTM has a tearing mode structure \cite{koslowski2005}. The distance of the PTM to coil PP804, slightly below the mid-plane on the outboard limiter, is approximately $0.19$~m.
\subsection{Helical structure and decay of the PTM}
\label{sec:structure_decay}
Typical application areas of the JET in-vessel magnetic coils are plasma control, equilibrium reconstruction and, with the fast arrays, the study of MHD phenomena just to name a few. Fast coil arrays are also used for MHD mode number analysis and hence to infer the periodicity of these signals. This was also done with PTM data by Koslowski et al. \cite{koslowski2005} in the past. They concluded that the PTM has a tearing mode structure and is formed in the vicinity of a resonant $q=3$ surface.\\

\begin{figure}[ht]
	\centering
	\includegraphics[width=15.00cm]{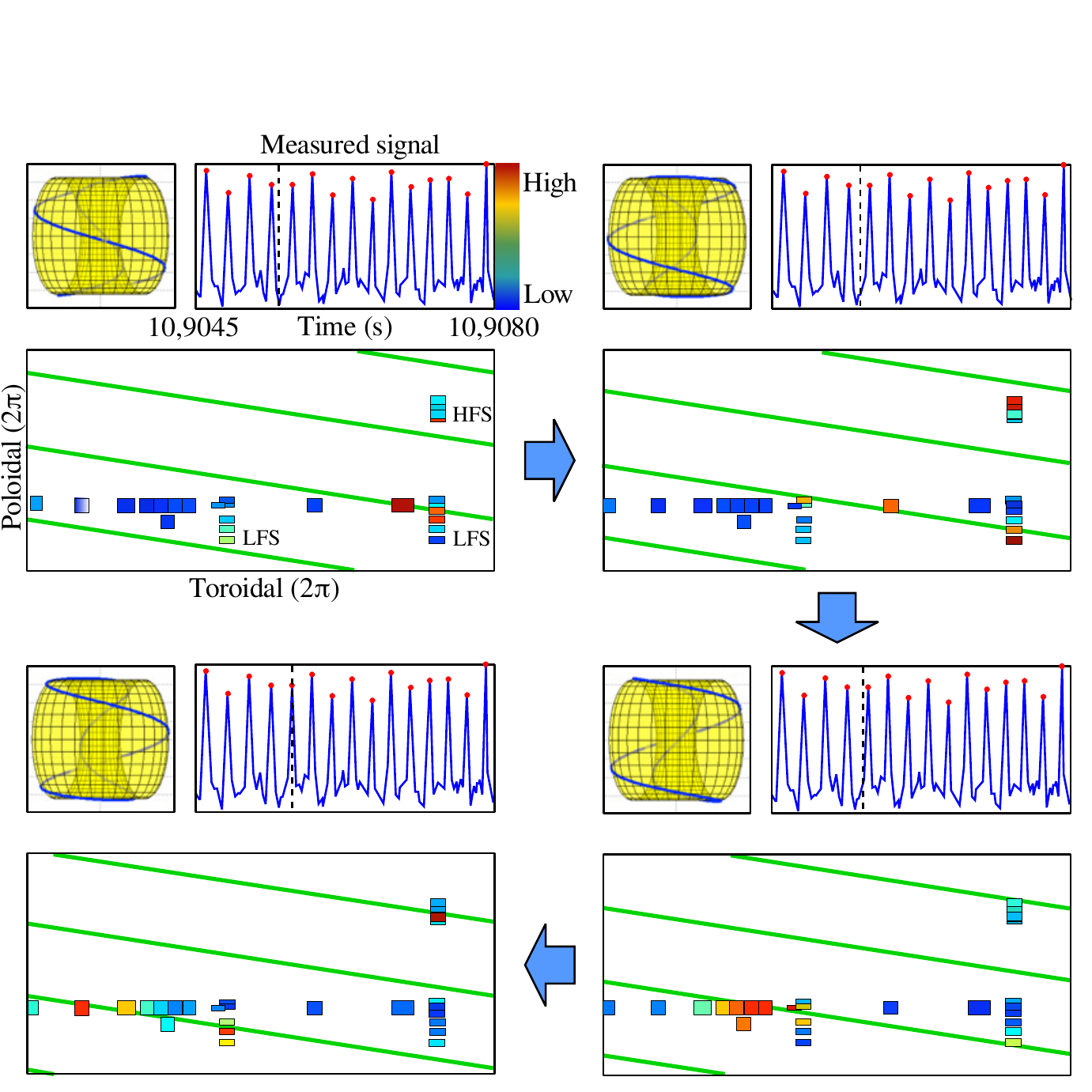}
	\caption{ The figure is based on a JET report of Benjamin Lagier \cite{ptm_lagier} and simplified for clarity.
		The figure shows four different time steps. Each plot is organized in three sections. In the upper left corner is the representation of a field line (blue) with constant $q=3$ on a rational surface for a torus with circular cross-section. The window in the upper right corner shows a measured signal from a fast coil in JET. Maxima are marked with red dots. The black vertical dashed line is a time marker. The lowest window represents the toroidal/poloidal plane. Limiter coil arrays on the high field side (HFS) and on the low field side (LFS) are indicated by HFS and LFS for easier orientation. Every used coil is represented as a rectangle where the measured signal strength is color-coded (red$\widehat{=}$high, blue$\widehat{=}$low signal strength). Green lines indicate the location of the simulated field line on a $q=3$ surface, synchronized with the measured signals (JET pulse: \#77188, $t=10.9045-10.9080$ ~s).}
	\label{fig:ptmpaperfig06}
\end{figure}

A complementary novel approach to infer the periodicity was performed by B. Lagier et al. \cite{ptm_lagier}. The available in-vessel magnetic coils were used to infer the helical structure by a comparison with a synthetic signal in the toroidal/poloidal plane. At first the magnetic signals were calibrated, numerically integrated and digitally filtered by a second order Butterworth high-pass filter with a cut-off frequency of 1~kHz. Afterwards the probe position were indicated in the toroidal/poloidal plane and the signal strength was colour-coded. If the perturbation of the PTM is close to one coil, the signal strength was high an therefore color-coded in red. Weak signals are color-coded blue. The results are presented in Fig.~\ref{fig:ptmpaperfig06}.\\

 The green lines in Fig.~\ref{fig:ptmpaperfig06} are a representation of a field line on a $q=3$ surface for a tokamak with circular cross-section. Once the simulated filament is synchronized with the experimental data, the highly localized simulated filament propagates in phase with the measured current perturbation of the PTM through the toroidal/poloidal plane. If the simulated filament is close to a coil also the measured signal is high. That is another evidence for the high localization of the PTM.\\

\begin{figure}[ht]
\centering
\includegraphics[width=15.00cm]{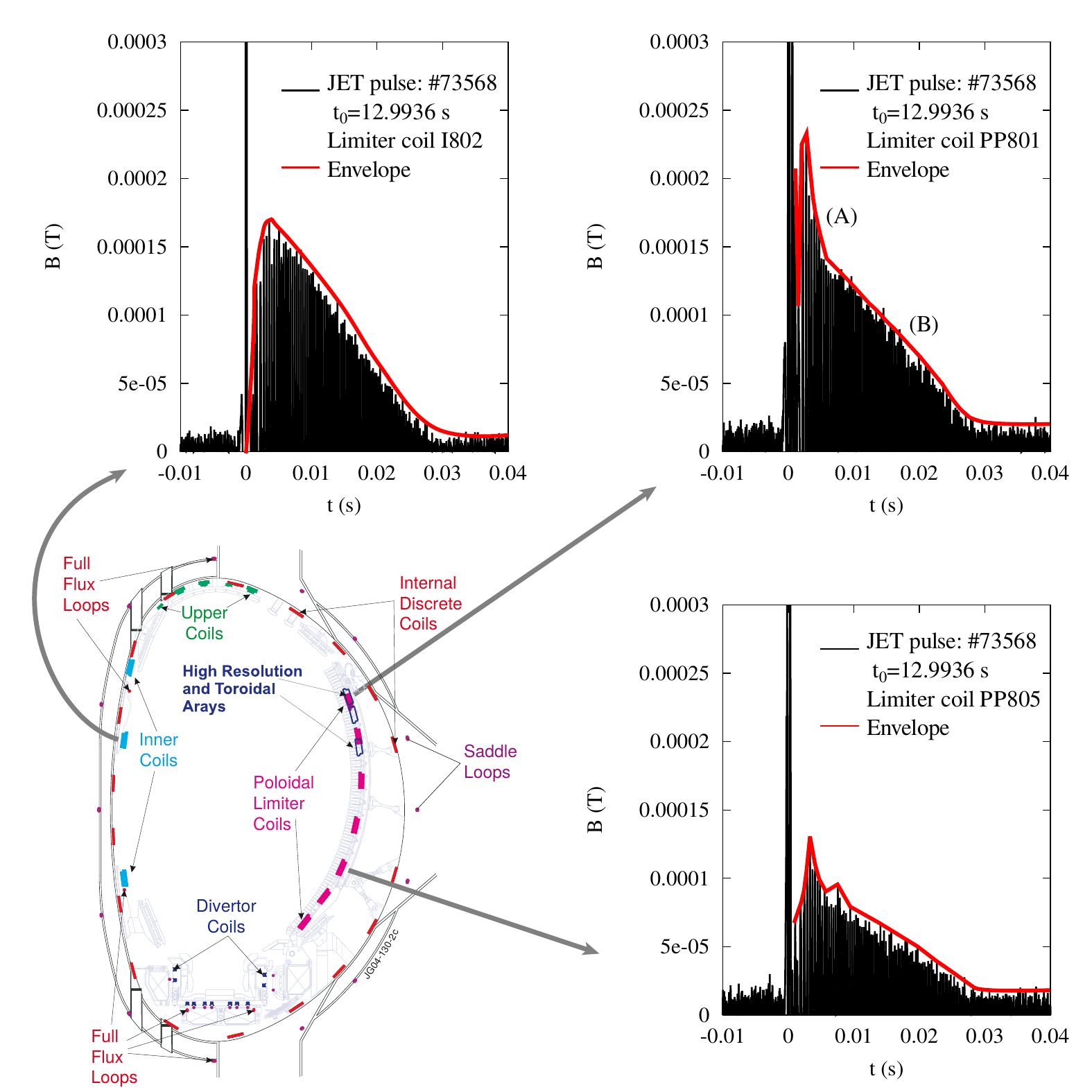}
\caption{Comparison of filtered and integrated magnetic signals at the outboard and inboard side of JET for three different coils (JET pulse: $\#73568$, $t_0=12.9936$~s). The envelopes of the signals are indicated with the red lines. (A) and (B) in the upper right panel indicate two time constants in signal decay.}
\label{fig:ptmpaperfig07}
\end{figure}

It is also instructive to compare the envelopes of the filtered and integrated PTM signals from magnetic coils at different positions around the torus in one poloidal plane (Fig.~\ref{fig:ptmpaperfig07}, JET pulse:  $\#73568$, $t_0=12.9936$~s).  At the inboard side (Fig.~\ref{fig:ptmpaperfig07}, coil I802) the mode grows for the first ms starting close at zero and decays after this phase almost linearly. Since ELM perturbations are more pronounced for the outboard side, the onset of the PTM can be traced back almost right after the ELM. The behaviour is different at the outboard side. Above the mid-plane (coil PP801) the onset of the mode is not as clear as at the inboard side since strong perturbations following the ELM-crash prevail. After this phase, two decay time constants can be identified. At first, the PTM shows a rapid decay phase (A) followed by a linear decay phase (B). Below the mid-plane (coil PP805) the amplitude rises quickly, peaks and after that process decays also with two time constants. In principle this could also be explained by a shift of  the whole plasma from the inboard to the outboard side. But the behaviour of the signal as measured by PP805 verifies why a movement of the whole plasma cannot cause this process. It would have to include also a rotation of the hole plasma in the electron diamagnetic drift direction to induce this signal. \\

Summarising, this spatio-temporal investigation of this PTM shows the best results because the PTM is located close to the in-vessel coils. PTMs are apparently highly localized structures (Fig.~\ref{fig:ptmpaperfig06}), which are co-rotating  with the bulk plasma (Sec.~\ref{sec:Rotation}) and show a temporal evolution that corroborates its generation at the outboard side (Fig.~\ref{fig:ptmpaperfig07}).
\section{Forward modelling of magnetic fluctuations}
From these experimental results we conclude that it should in principle be possible to describe the PTM phenomenology as rotating closed current filament in the pedestal region where an initially localised current distribution is allowed to expand parallel to the magnetic field on a rational surface.\\
Following the approach in \cite{migliucci2010}, Biot-Savart's law was used to compute the magnetic field at different coil positions and times. The helical structure of the filament was obtained for a single field line on a rational $q=3$ surface with FLUx Surface Handling (FLUSH) \cite{flush}. It is a collection of FORTRAN routines to assist in reconstructing and post-processing the flux solution of a plasma equilibrium solver at JET. To be more specific in this study the subroutines FLUQAX and FLUPN3 were used for JET pulse: $\#73568$ at $t=12.98$~s. The angular velocity of the filament was mimicked using Eq.~\ref{equ:ptmrot}. The distance of the mode $d\approx0.19$~m to the pick-up coil PP804 was estimated using the radial position information from the ECE diagnostic (Fig.~\ref{fig:ptmpaperfig04}).\\
As initial condition a Gaussian shaped current density distribution along the parallel direction of the field line with a maximum current density of $j_0=4.2\times 10^5$~A/m$^2$, standard deviation $\sigma=$1.2~m and a filament cross-section of $A_0=1.8\times 10^{-3}$~m$^2$ was placed in the middle of the domain at $t_0$. This peak current density - giving a peak magnetic field strength - was matched with the maximum field for coil PP801 in Fig.~\ref{fig:ptmpaperfig07}. \\

In order to limit the lifetime of the modelled filament to $t_{lt}=25$~ms a dissipation mechanism is required. In this study we treat the filament as a short solenoid with $N=3.0$ windings and a series resistance $R$. The resistance is determined via the parallel Spitzer resistivity $\eta_\parallel$ and the length of a filament $l_f$ on a $q=3$. In the case of JET this is approximately $l_f=56$~m. The electron temperature at the location of the PTM after the collapse of the pedestal is roughly $T_e\approx280$~eV. The parallel Spitzer resistivity is therefore around $\eta_\parallel=7.0\times 10^{-8}$~$\Omega$m and the Ohmic resistance therefore $R=3.9\times 10^{-6}$~$\Omega$.\\
The inductance $L$ of a short solenoid is computed according to \cite{wheeler1928}
\begin{equation}
L(\mu_r)\approx \mu_0\mu_r\frac{N^2A}{h_f+0.8r_f}
\end{equation}
where $h_f=3$~m is the length, $r_f=4$~m the radius of the solenoid and $A=\pi r_f^2$ the area of the cross-section. The time constant for the current decay is $\tau=L(\mu_r)/R$. We assume that the current is gone after five time constants $\tau=t_{lt}/5=0.005\,$s. With $R=3.9\times 10^{-6}$~$\Omega$ and $\tau=0.005$~s we compute for the inductance $L$ and the relative permeability $\mu_r$
\begin{equation}
L=1.98\times 10^{-8}\,\textrm{H} \qquad \mu_r=0.23\times 10^{-3}.
\end{equation}

\begin{figure}[ht]
\centering
\includegraphics[width=15.00cm]{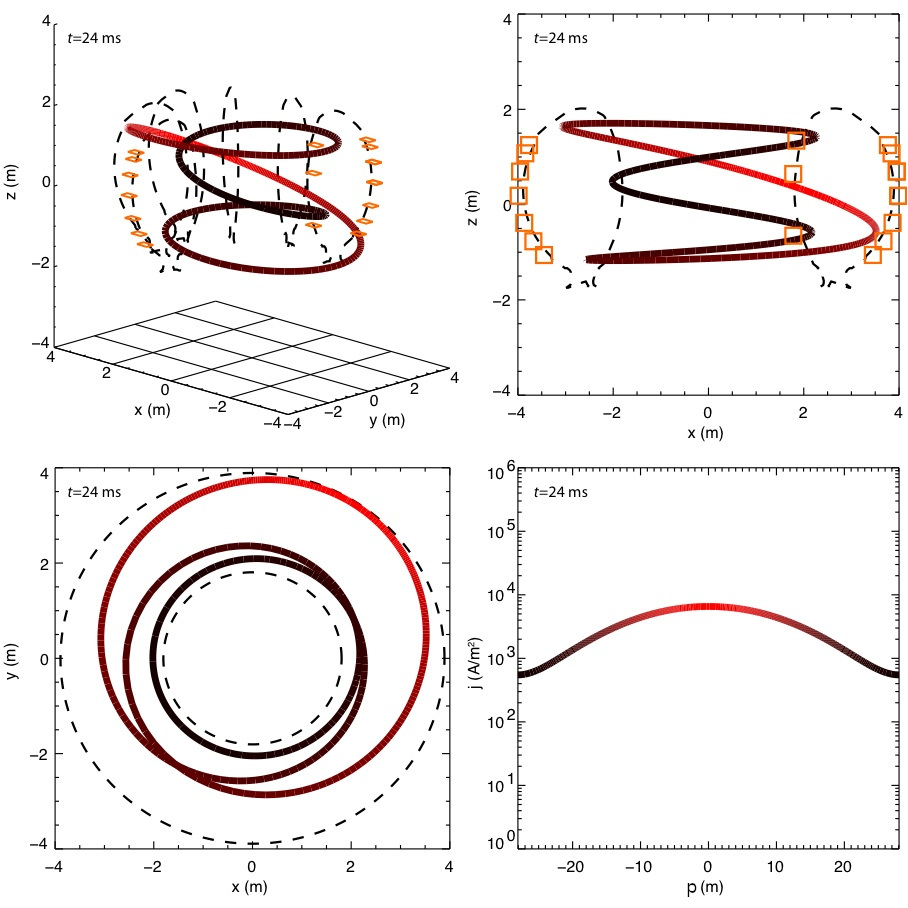}
\caption{Illustration of a filament with locally varying $q$ on a $q=3$ surface, calculated with FLUSH in various presentations. The lower right panel shows the current density $j$ of the filament in parallel direction $p$. The colored squares in the two upper panels represent the magnetic pick-up coils at the limiters in the octants four and eight. The current distribution after $24$~ms in the filament is color-coded. Red means high current density, black low current density. For the exact values see the lower right panel please (Equilibrium from JET pulse: $\#73568$, $t=12.98$~s).}
\label{fig:ptmpaperfig08}
\end{figure}

For the numerical simulation with a time resolution of $1\times 10^{-7}$~s, periodic boundaries where chosen to reflect the nature of a closed current filament (Fig.~\ref{fig:ptmpaperfig08}).\\
\subsection{Results}
\begin{figure}[ht]
\centering
\includegraphics[height=7.00cm]{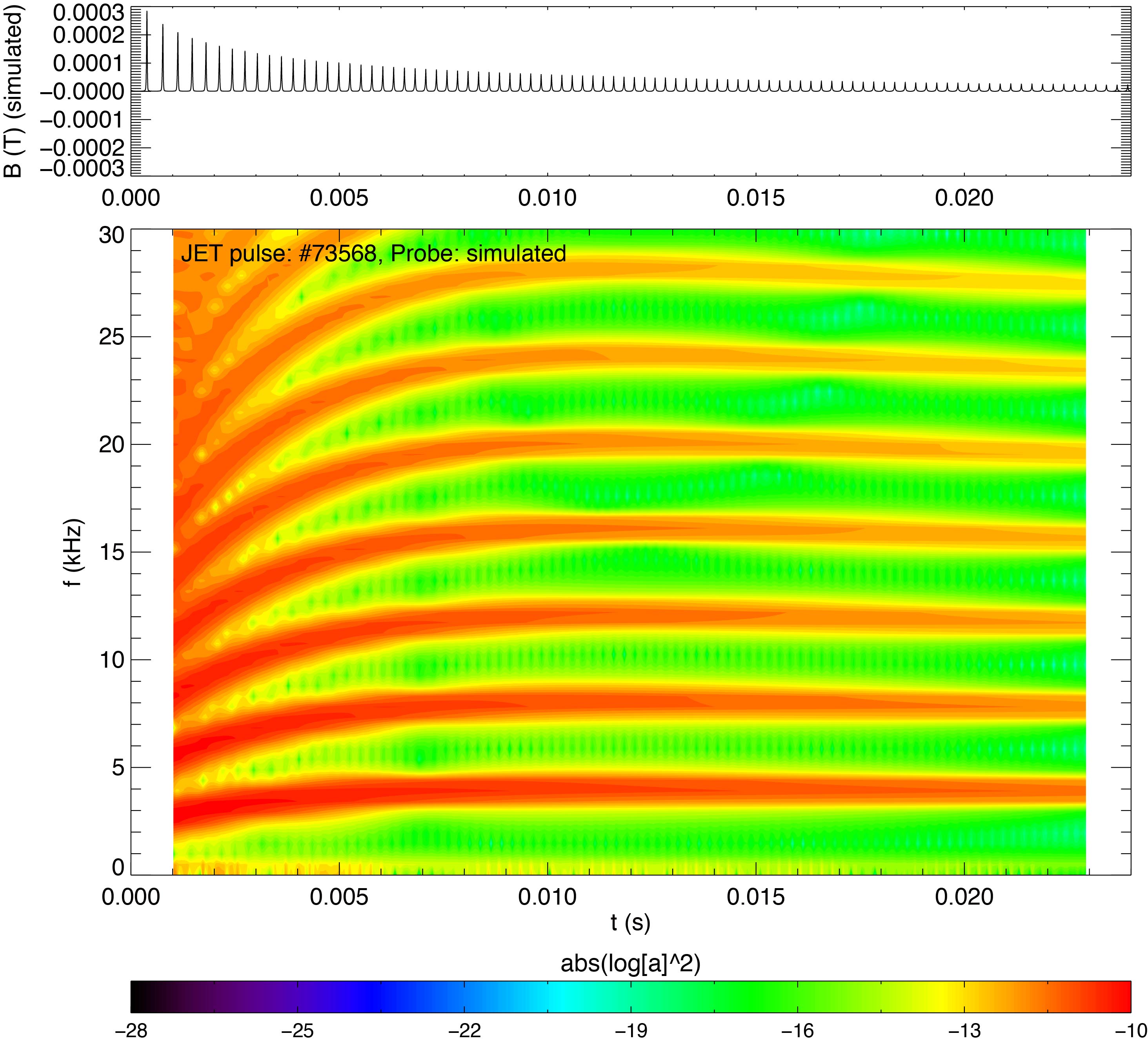}
\caption{FFT of magnetic fluctuations from a forward-modeled closed rotating current filament.}
\label{fig:ptmpaperfig09}
\end{figure}

A comparison of FFTs from measured and integrated data (Fig.~\ref{fig:ptmpaperfig01}) and from forward-modeled magnetic fluctuations (Fig.~\ref{fig:ptmpaperfig09}) show good agreement. Both plots show the characteristic harmonics, a sharp increase of the frequency in the beginning and a slow decay of the mode after approximately $25$~ms. Obviously, a highly localized current filament is also capable to produce the observed spectra.\\

Forward modelling of magnetic fluctuations gives information both on the direction of the current and the periodicity of the current filament. The current is approximately $I\approx-294.5$~A at $t=12.996$~s. The current density is then of the order of $-0.15$~MA/m$^2$ assuming a filament cross-section of $A_0=1.8 \times 10^{-3}$~m$^2$ \cite{koslowski2005}. The polarity (sign) of the coils was defined using the disruptions $\#73460$, $\#77176$, the $m/n=2/1$ mode in $\#77635$  ($t=44.207-44.2084$~s) and the dry run $\#77638$  ($40-43$~s. P4 blip window) by Sergei Gerasimov et al\footnote{Private communication, \cite{Gerasimov2010}}. The sign of the current indicates a lack of current compared to the plasma current and therefore a current hole in the PTM region.\\

\begin{figure}[ht]
\begin{minipage}[b]{0.45\linewidth}
\centering
\includegraphics[height=7cm]{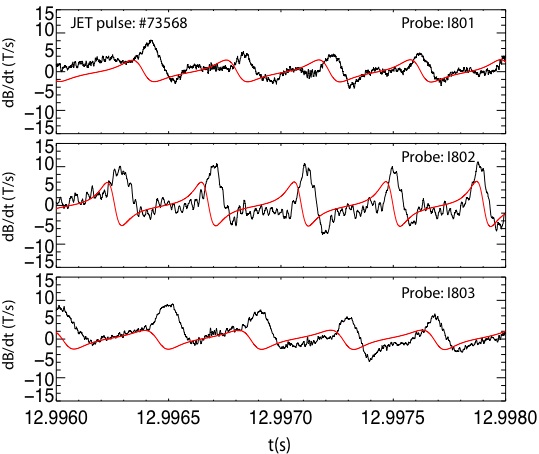}
\caption{Comparison of a simulated unipolar current filament (red) with signals from the inboard array (black, I801-I803) for JET pulse $\#73568$.}
\label{fig:ptmpaperfig10}
\end{minipage}
\hspace{0.5cm}
\begin{minipage}[b]{0.45\linewidth}
\centering
\includegraphics[height=7cm]{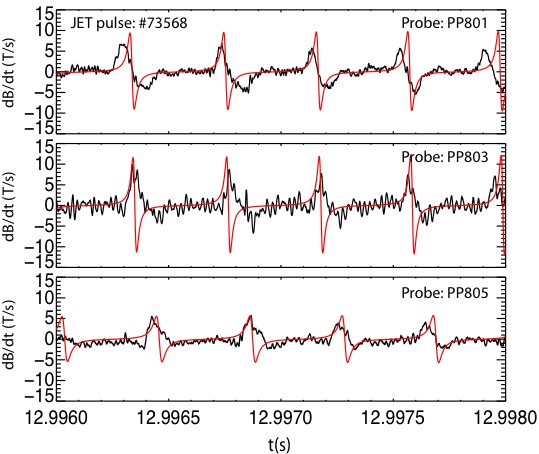}
\caption{Comparison of a simulated unipolar current filament (red) with signals from the outboard array (black, PP801-PP805) for JET pulse $\#73568$.}
\label{fig:ptmpaperfig11}
\end{minipage}
\end{figure}

The signals of magnetic signatures from an unipolar current filament are comparable to the envelopes of the measured signals (Fig.~\ref{fig:ptmpaperfig10} and Fig.~\ref{fig:ptmpaperfig11}). For the coils PP801, PP803, PP805 the signal strength is overestimated. Although fairly well synchronized with PP801, there is a small phase shift between the signals in PP803 and PP805. This is even worse for signals at the inboard side where the current is underestimated. Nevertheless the shape of the signals approve the negative sign. Errors in the phase relation are probably due to use of a poor equilibrium where the edge current and distortions of the PTM are not reflected.\\

The presented model is simplified, therefore we expect only qualitative agreement with the observations.
\section{Discussion}
Based on the results from the experiment an intuitive model is proposed which could in principle explain the genesis of the PTM.\\

\begin{figure}[ht]
\centering
\includegraphics[height=7cm]{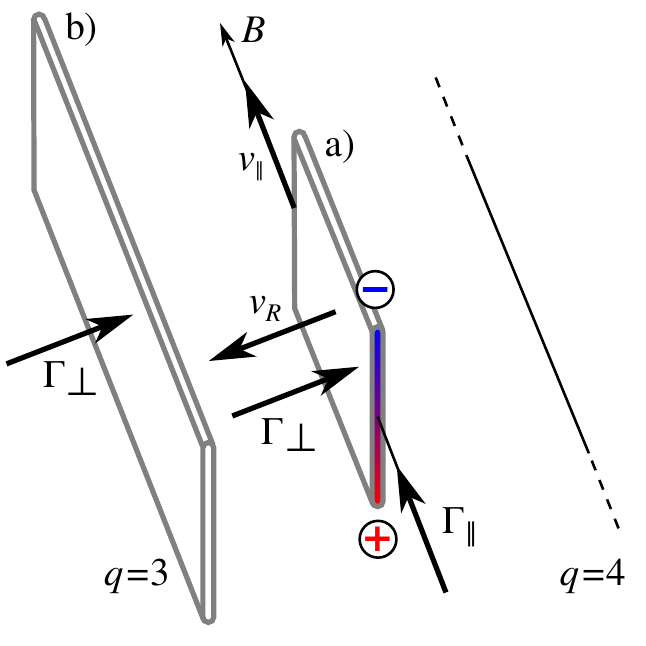}
\caption{a) A hole is generated between the $q=3$ and $q=4$ surfaces. The localized, $B$ field-aligned, polarized hole travels up the density gradient with the radial advection velocity $v_R$ and expands in parallel direction with velocity $v_\parallel$. In this stage, it is filled by parallel $\Gamma_\parallel$ and perpendicular transport $\Gamma_\perp$.\\
b) The hole was able to reach a field line on the $q=3$ surface before it was completely filled and closed on itself. $v_R=0$ since the charge imbalance is short-circuited by parallel currents. It is now filled by perpendicular transport $\Gamma_\perp$ only.
}
\label{fig:ptmpaperfig12}
\end{figure}

Langmuir probe measurements in JET reveal that blobs and holes are created in the edge shear layer. Whereas blobs can propagate large distances and in principle can reach the first wall, holes travel up the temperature gradient and are only observed slightly inside the region where they have been born. Calculations show that holes are filled on the conductive time scale $\tau_{cond}$. Since $\tau_{cond}\approx L_\parallel/\chi_{e\parallel}$ depends on the parallel length $L_\parallel$ and the electron thermal conductivity $\chi_{e\parallel}$, the lifetime of holes is in the range of $\mu$s \cite{xu2009}. The lifetime of a hole is therefore influenced by its size and the electron temperature. Additionally, low shear rates are a prerequisite. Otherwise holes are sheared apart rather quick.\\
A localized, $B$ field aligned hole wich is travelling to a resonant magnetic surface with a radial velocity $v_R$ exhibits a current, a density and a temperature perturbation (Fig.~\ref{fig:ptmpaperfig12}a). It is expanding parallel to the field line with velocity $v_\parallel$. Holes are usually quickly filled by parallel $\Gamma_\parallel$ and perpendicular transport $\Gamma_\perp$. Fig.~\ref{fig:ptmpaperfig12}b) shows the situation if such a hole was able to reach a field line on a resonant surface. Then, the polarization is short-circuited by parallel currents and the motion stops. The hole is then filled by perpendicular transport only which increases its lifetime significantly.\\

\begin{figure}[ht]
\centering
\includegraphics[height=7cm]{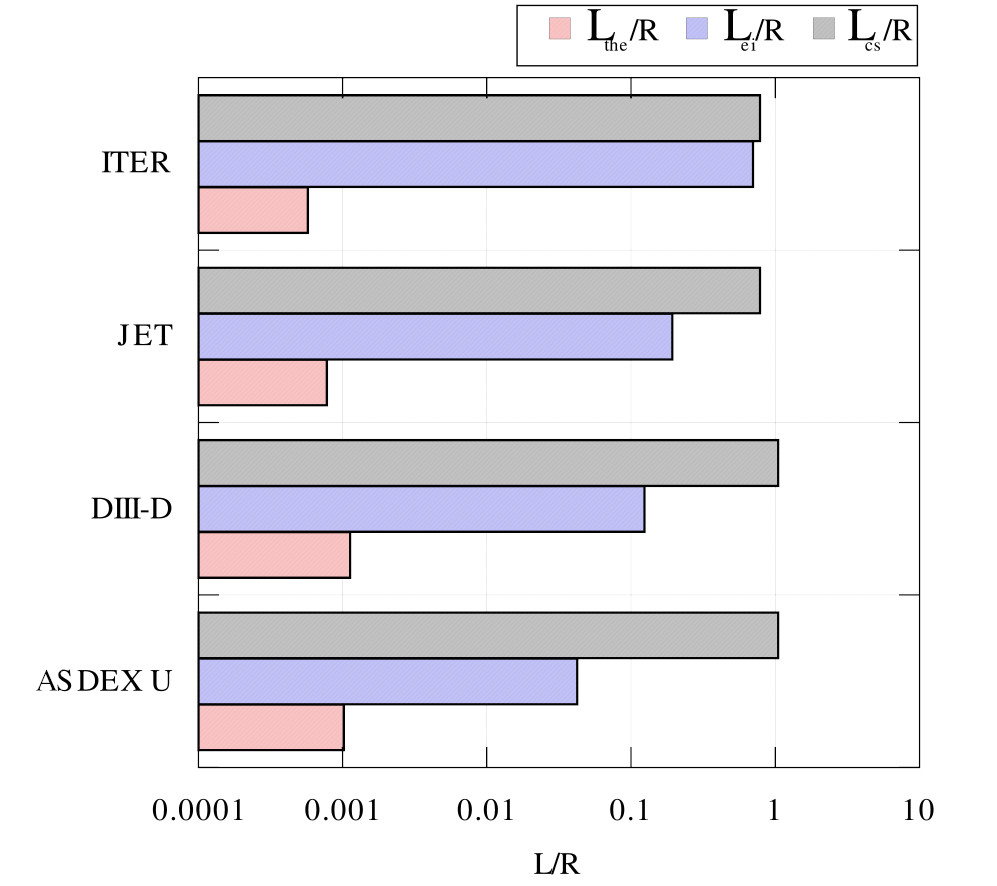}
\caption{Multimachine comparison of maximum radial propagation distances of holes under the assumption that these holes are filled on density $\tau_{cs}$, thermal $\tau_{the}$ and resistive $\tau_{ei}$ time scales. All corresponding distances ($L_{cs}$, $L_{the}$ and $L_{ei}$) are normalized by the major radius $R$ of the respective machine.}
\label{fig:ptmpaperfig13}
\end{figure}

Following that line of thought, plasma edge properties should have a large impact on the lifetime and whether a PTM can be observed or not. In the following section, a simple multi-machine comparison is presented for the tokamaks ASDEX Upgrade, DIII-D, JET and ITER (Fig.~\ref{fig:ptmpaperfig13}).\\

The radial advection velocity $v_R$ of the hole is assumed to be caused by the nonlinear evolution of interchange motions, without any presumption of plasma sheaths. Radial velocities and radial extensions are computed according to the model in \cite{garcia2005}. Sheath dissipative models like in \cite{krasheninnikov2001} are not appropriate in this case because of the high radial velocities of the structures. The hole enters and leaves any given flux tube so quickly that the plasma in the flux tube cannot react to the pressure perturbation \cite{fundamenski2007}. The particle density and the edge safety factor for ASDEX Upgrade, JET and DIII-D were taken from \cite{maggi2007}. The associated pedestal temperatures for H-mode plasma with type-I ELMs were selected from \cite{hatae2001}. The required information for ITER was found in \cite{ITER_sugihara} and \cite{snyder2009}.\\
 Figure~\ref{fig:ptmpaperfig13} summarizes the maximum normalized propagation distances depending on the filling mechanism. A hole with parallel length $L_{blob}$ \cite{gunn2007} and radial velocity $v_R$ can travel the distance $L_{cs}$ if it is filled by ions with sound speed $c_s$. Similarly, distances can be calculated if the hole is filled by thermal electrons (distance $L_{the}$) or if the current decays according to the Spitzer resistivity (distance $L_{ei}$). The raw data and calculations are summarized in the appendix in Tab.~\ref{tab:multimachine}.\\
Holes can travel slightly further in ASDEX Upgrade and DIII-D on thermal $\tau_{the}$ and density time scales $\tau_{cs}$ if compared to JET or ITER. This is due to the lower pedestal temperatures for the smaller machines. The strongest  influence on the propagation distance is found on the resistive time scale $\tau_{ei}$. Figure~\ref{fig:ptmpaperfig13} shows that the collisionality in the pedestal has the strongest influence on the lifetime of holes. Therefore, ITER is expected to excel when it comes to lifetimes and propagation distances of holes.\\

\begin{figure}[ht]
\centering
\includegraphics[height=8.00cm]{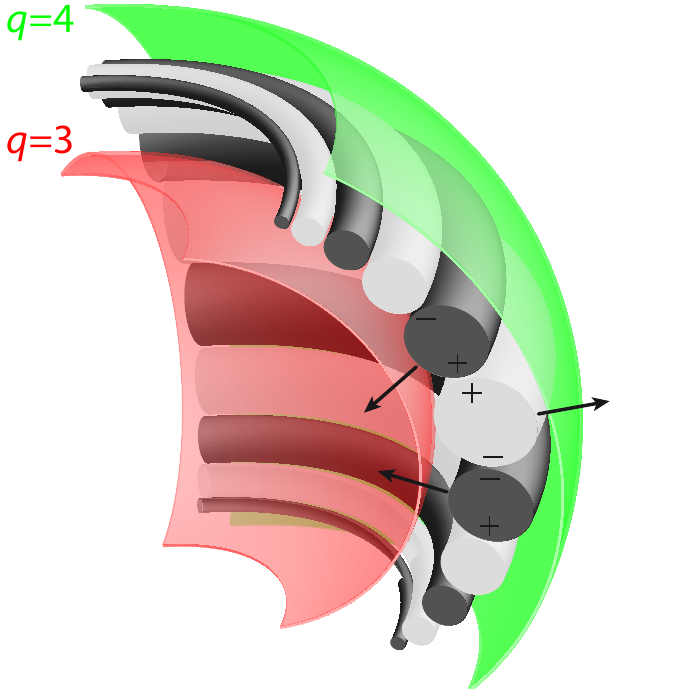}
\caption{Illustration of an ELM in the exhaust phase at the outboard side of a tokamak. If an ELM filament (white) is stopped and captured at e.g. the $q=4$ surface (green) an outer mode or current ribbon could be created. Similarly a hole (black) captured on a $q=3$ surface could create a PTM.}
\label{fig:ptmpaperfig14}
\end{figure}

From pellet injection experiments in DIII-D and Tore Supra it is known that the maxima of the mass deposition profiles are close to the magnetic $q=2$ and $q=3$ surfaces. Low order rational surfaces appear to slow down or even stop the polarization drift of the density cloud towards the low field side of the plasma \cite{commaux2010}. Since holes are regions of lower density than the ambient plasma the $\nabla B$ polarization and therefore the drift is in opposite direction to blobs or density clouds. Additionally, it is similarly to expect that holes can be slowed down or trapped in the vicinity of a rational surface. Under the assumption that the interchange drive of a polarized ELM-induced hole moves it into the proximity of a rational surface with low $q$, it could be slowed down or even stopped. An illustration of an ELM in the exhaust phase is given in Fig.~\ref{fig:ptmpaperfig14}. Holes are indicated as black filaments, ELM filaments are in white. Large ELM induced holes could close on themselves in the vicinity of a rational magnetic field line and therefore detach from the magnetic surface. During this process a closed filament would be created. For a closed filament, charge interchange imbalance would be short-circuited by parallel currents which would stop the filament completely. The result would be a closed, well localized filament with a significantly increased lifetime because it could be only filled by slow-perpendicular transport. This filament should then have a distinct and rational ratio of toroidal and poloidal periodicity. The lack of current inside the filament would change the equilibrium locally there. \\

In our opinion a promising candidate for the described events above is the Palm Tree Mode (PTM), an ELM post-cursor. As presented, forward-modeled magnetic fluctuations from a current filament are in a fair agreement with experimental results. The physical properties of PTMs are compatible with the properties expected for holes after ELM-filaments are generated. The strong influence of the collisionality in the pedestal on the lifetime of the holes explain why PTMs are only observed in JET up to now (Fig.\ref{fig:ptmpaperfig13}). For ITER it is expected that these phenomena will also be present and influence the confinement in the edge region.\\
A candidate for trapped blobs are outer modes which are current filaments with excess current on a rational surface \cite{solano2010}. Taking the proposed model into account outer modes are ELM filaments, which were not able to propagate to the scrape-off layer (SOL). If present, outer modes mitigate ELMs.  In our view, PTMs are remnants of ELM-induced holes. Both of these structures are born and trapped in the vicinity of rational surfaces. Nevertheless, the probability to create PTMs should be lower compared to outer modes because ELM induced holes would have to travel up the temperature gradient into regions with increasing parallel transport and could be filled before they can be trapped on a rational surface.
\section{Conclusion}
This paper presented how the concept of blob and hole-like transport can contribute to shed more light on the genesis and nature of Palm Tree Modes and Outer modes in the JET tokamak. It was demonstrated that under the current filament assumption it is possible to forward-model the magnetic signatures of PTMs which shows the same harmonics structure and reflect also the signal shapes reasonably well.\\
Additionally, it was demonstrated how the JET in-vessel magnetic coils can be utilized to get more insights in the genesis and nature of PTMs.\\
In the introduction in Sec.~\ref{sec:introduction} we raised the question whether ELMs create current density and density holes in the edge plasma. Based on our findings we suggest that PTMs are trapped current and density holes produced by ELMs. A fruitful further direction of research could be whether or not L-mode blobs and holes in ITER could be big enough to produce similar structures. This could have implications on the L-mode confinement and the LH-transition.\\
We also conclude that the rich harmonics in FFTs are probably not due to MHD-modes but rather because of the high localisation of the filament. FFT spectra are therefore probably misleading in PTM analysis. Other tools like empirical mode decomposition and Hilbert amplitude spectra therefore give a better representation of the physical nature of the PTM \cite{maszl2011b}.
\section{Acknowledgements} This work has been carried out within the framework of the EUROfusion Consortium and has received funding from the EURATOM research and training programme 2014-2018 under Grant Agreement No. 633053. The views and opinions expressed herein do not necessarily reflect those of the European Commission. The support by the Austrian Science Fund (Fonds zur Förderung der wissenschaftlichen Forschung in Österreich) under grant No. P19901 is also gratefully acknowledged. Additional support was lend by the österreichische Forschungsgemeinschaft (Austrian Research Community) under grant 06/11454, International Communication.\\
One of the authors, Christian Maszl, wants to thank Codrina Ionita, Thijs W. Versloot, Benjamin Lagier and Mathias Brix for valuable discussions and support during his visits at JET and the DTU plasma physics group for their hospitality.
\section*{References}
\bibliography{bibliothek}
\bibliographystyle{unsrt}
\clearpage
\section*{Palm tree mode database}
PTMs where searched manually with the IDL program SPECVIEW\index{SPECVIEW} at the JET Analysis Cluster (JAC)\index{JAC}\index{JET analysis cluster}. Tab.~\ref{tab:ptm_db_I} and Tab.~\ref{tab:ptm_db_II} summarizes the palm tree modes found. The second column gives the time windows where the PTMs appear. The abbreviation TBE (terminated by ELM) indicates if the mode was stopped by a subsequent ELM (y) or reached its unperturbed lifespan (n). The hyphen (-) marks inconclusive cases. TW marks pulses were no fast time window is available. The columns ECE and MAGS give the sample frequencies of the respective diagnostics. The MHD section documents the fundamental frequencies of other MHD activity. The year column allows a discrimination between the different divertor configurations (Tab.~\ref{tab:divertor}).\\
\begin{table}[ht]
\caption{Divertor configurations in JET from 1999-2009.}
\begin{center}
\begin{tabular}{|l|l|l|}
Period		&	Name						&	Abbrevation	\\
\hline
1999-2001 	&	MkII Gas Box					&	MkII GB		\\ 
2002-2004 	&	MkII Septum Replacement Plate 			&	MkII SRP 	\\
2005-2009 	&	MkII High Delta 				&	MkII HD		\\
\hline
\end{tabular}
\end{center}
\label{tab:divertor}
\end{table}%

\begin{table}[ht]
\caption{Palm tree mode database - part I}
\begin{center}
\begin{tabular}{|c|c|c|c|c|c|c|}
\hline
Pulse     & t (s)                   & TBE & ECE           & MAGS        & MHD        &	Year		\\
\hline
52011   &   59.14-59.22   &   -    &   250 kHz   &   250 kHz   &   39 kHz  & 	2000		\\
58982   &   57.26-57.39   &   y   &   250 kHz   &    250 kHz  &   12 kHz  &	2003		\\
72299   &   61.44-61.48   &   n   &   TW    	     &       1 MHz  &                 &	2008		\\
73568   &   52.80-53.20   &   n   &   250 kHz   &   2 MHz      &   39 kHz  &	2008		\\
73569   &   51.37-51.44   &   n   &    TW               &   2 MHz      &   29 kHz  &	2008		\\
74812   &   54.56-54.64   &   -    &       TW              &   1 MHz      &   11 kHz  &	2008		\\
74812   &   56.35-56.45   &   n   &             TW        &   1 MHz      &                  &	2008		\\
75411   &   57.10-57.20   &   n   &               TW      &   1 MHz      &                  &	2008		\\
75411   &   57.86-57.92   &   n   &       TW              &   1 MHz      &                  &	2008		\\
76793   &   49.60-49.80   &   -    &             TW        &    1 MHz                 &   9 kHz    &	2009		\\
77184   &   53.35-53.40   &   -    &                 TW    &   1 MHz      &   32 kHz  &	2009		\\
77185   &   55.75-55.95   &   n   &             TW        &   2 MHz      &                  &	2009		\\
77186   &   51.12-51.15   &   y   &                  TW   &   1 MHz      &                  &	2009		\\
77188   &   50.89-50.93   &   y   &                 TW    &   1 MHz      &                  &	2009		\\
77319   &   57.10-57.15   &   y   &                  TW   &   1 MHz      &                  &	2009		\\
77326   &   63.90-64.05   &   n   &                 TW    &   1 MHz      &                  &	2009		\\
77326   &   65.10-65.20   &   n   &                  TW   &   1 MHz      &                  &	2009		\\
77326   &   65.50-65.70   &   n   &                   TW  &   1 MHz      &                  &	2009		\\
\hline
\end{tabular}
\end{center}
\label{tab:ptm_db_I}
\end{table}%

\begin{table}[ht]
\caption{Palm tree mode database - part II}
\begin{center}
\begin{tabular}{|c|c|c|c|c|c|c|}
\hline
Pulse     & t (s)                   & TBE & ECE           & MAGS        & MHD        &	Year		\\
\hline
77326   &   66.10-66.20   &   n   &   TW    	     &  1 MHz      &                  &	2009		\\
77329   &   64.20-64.50   &   n   &   TW    	     &   1 MHz      &                  &	2009		\\
77329   &   65.30-65.40   &   n  &   TW    	     &   1 MHz      &                  &	2009		\\
77329   &   65.90-66.00   &   n   &   TW    	     &1 MHz      &                  &	2009		\\
77331   &   65.30-65.40   &   n  &   TW    	     &   1 MHz      &                  &	2009		\\
77332   &   66.00-66.12   &   y   &   TW    	     &   1 MHz      &                  &	2009		\\
77335   &   63.90-64.10   &   n   &   TW    	     &   1 MHz      &                 &	2009		\\
77335   &   65.30-65.45   &   n   &   TW    	     &   1 MHz      &                 &	2009		\\
77342   &   63.90-64.00   &   n   &   TW    	     &   1 MHz      &                 &	2009		\\
77342   &   64.34-64.41   &   n   &   TW    	     &   1 MHz      &                 &	2009		\\
78690   &   55.40-55.50   &   n   &   TW    	     &   1 MHz      &                 &	2009		\\
78712   &   53.10-53.60   &   n   &   TW    	     &   1 MHz      &   11 kHz  &	2009		\\
78712   &   55.40-55.70   &   n   &   TW    	     &   1 MHz      &                  &	2009		\\
79619   &   57.25-57.35   &   n   &   TW    	     &   1 MHz      &   13 kHz  &	2009		\\
79620   &   45.85-45.92   &   n  &   TW    	     &   1 MHz      &                  &	2009		\\
79620   &   46.49-46.58   &   n   &   TW    	     &   1 MHz      &                  &	2009		\\
79744   &   58.10-58.25   &   n   &   TW    	     &   1 MHz      &   16 kHz   &	2009		\\
79745   &   61.10-61.30   &   n   &   TW    	     &    1 MHz     &   26 kHz   &	2009		\\
\hline
\end{tabular}
\end{center}
\label{tab:ptm_db_II}
\end{table}%
\clearpage
\section*{Multi-machine comparison}
\begin{table}[ht]
	\centering
	\begin{tabular}{|l|c|c|c|c|l|}
\textbf{Tokamak} & \textbf{JET} & \textbf{ASDEX U} & \textbf{ITER} & \textbf{DIII-D} & \\
\hline
$n$ (m$^{-3}$)	& $6.0\times 10^{19}$	& $6.0\times 10^{19}$	& $1.0\times 10^{20}$	& $6.0\times 10^{19}$	& particle density\\
$T_{e,i}$ (eV)	& 1200			& 400			& 5000			& 700			&
temperatures \\
$\ln \Lambda$	& 16			& 15			& 17			& 15			& Coulomb log. \\
$\eta_\parallel$ ($\Omega$m) & $2.0\times10^{-8}$ & $9.8\times10^{-8}$ & $2.5\times10^{-9}$ & $4.8\times10^{-8}$ & Spitzer res.\\
$R$ (m)		& 3.1			& 1.65			& 6.2			& 1.69			& major radius \\
$q$		& 3.0			& 4.0			& 3.0			& 4.0			& safety factor\\
$B$ (T)		& 2.3			& 3.0			& 5.0			& 3.0			& magnetic field\\
$\omega_{ci}$ (Hz) & $1.1\times 10^8$	& $1.5\times 10^8$	& $2.4\times 10^8$	& $1.5\times 10^8$	& $i^+$ gyro freq.\\
$c_s$ (m/s)	& $3.4\times 10^5$	& $2.0\times 10^6$	& $6.9\times 10^5$	& $2.6\times 10^5$	& sound speed \\
$v_{the}$ (m/s) & $2.1\times 10^7$	& $1.2\times 10^7$    	& $4.2\times 10^7$	& $1.6\times 10^7$	& $e^-$ thermal vel.\\
$\rho_s$ (m)	& $3.1\times 10^{-3}$	& $1.4\times 10^{-3}$	& $2.9\times 10^{-3}$	& $1.8\times 10^{-3}$	& $i^+$ gyro rad.\\
\hline
\multicolumn{6}{l}{\textbf{Blob}} \\
\hline
$L_{blob}$ (m)	& 2.4			& 1.7			& 4.9			& 1.8			& parallel length \\
$\Delta n/n$	& 0.5			& 0.5			& 0.5			& 0.5 			& perturbation\\
$R_{blob}$ (m)	& $1.1\times 10^{-2}$	& $5.7\times 10^{-3}$	& $1.2\times 10^{-2}$	& $7.2\times 10^{-3}$	& radial blob size \\
$v_{blob}$ (m/s)& $2.0\times 10^4$	& $1.2\times 10^4$	& $3.1\times 10^4$	& $1.7\times 10^4$	& blob velocity \\
\hline
\multicolumn{6}{l}{\textbf{Timescales}}\\
\hline
$\tau_{cs}$ (s)	& $7.2\times 10^{-6}$	& $8.8\times 10^{-6}$	& $7.0\times 10^{-6}$	& $6.8\times 10^{-6}$	& $L_{blob}/c_s$\\
$\tau_{the}$ (s)& $1.2\times 10^{-7}$	& $1.5\times 10^{-7}$	& $1.2\times 10^{-7}$	& $1.1\times 10^{-7}$	& $L_{blob}/v_{the}$\\
$\tau_{ei}$ (s) & $3.0\times 10^{-4}$	& $6.1\times 10^{-4}$	& $1.4\times 10^{-4}$	& $1.2\times 10^{-5}$	& $ne^2/m \eta_\parallel$\\
\hline
\multicolumn{6}{l}{\textbf{Propagation}}\\
\hline
$L_{cs}$ (m)	& 2.4			& 1.7			& 4.9			& 1.8			& $v_{blob}\cdot\tau_{cs}$\\
$L_{the}$ (m)	& $2.4\times 10^{-3}$	& $1.7\times 10^{-3}$	& $3.6\times 10^{-3}$	& $1.9\times 10^{-3}$	& $v_{blob}\cdot\tau_{the}$\\
$L_{ei}$ (m)	& 0.6			& $7.0\times 10^{-2}$	& 4.4			& 0.2			& $v_{blob}\cdot\tau_{ei}$\\
\hline
\end{tabular}
	\caption{Multimachine comparison for characteristic blob sizes, timescales and propagation estimates.}
	\label{tab:multimachine}
\end{table}
\end{document}